\documentstyle[12pt]{article}
 \def\be{\begin{eqnarray}}
 
 \def\ee{\end{eqnarray}}
\renewcommand{\textfraction} {.01} 
\renewcommand{\topfraction} {.99} 
\input psfig.sty
\begin{document} 
\pagestyle{empty}
\Huge{\noindent{Istituto\\Nazionale\\Fisica\\Nucleare}}

\vspace{-3.9cm}

\Large{\rightline{Sezione di ROMA}}
\normalsize{}
\rightline{Piazzale Aldo  Moro, 2}
\rightline{I-00185 Roma, Italy}

\vspace{0.65cm}

\rightline{INFN-1321/01}
\rightline{July 2001}

\vspace{1.cm}

\renewcommand{\textfraction} {.01} 
\renewcommand{\topfraction} {.99} 
 
\begin{center}{\large\bf  Neutron structure function $F_2^n(x)$ 
from deep inelastic
electron scattering off few-nucleon systems   }
\end{center}

\begin{center} E. Pace$^a$, G. Salm\`{e}$^b$,  S. Scopetta$^c$ 
and  A. Kievsky$^d$
\end{center}

\noindent { $^a$ \it Dipartimento di Fisica, Universit\`{a} di
Roma "Tor Vergata" and Istituto Nazionale di   Fisica Nucleare, Sezione Tor
Vergata, Via della Ricerca Scientifica 1, I-00133 Roma,  Italy}

\noindent { $^b$ \it Istituto Nazionale di Fisica Nucleare, Sezione
di  Roma, P.le A. Moro 2, I-00185 Rome, Italy}

\noindent { $^c$ \it  Dipartimento di Fisica,
Universit\`a degli Studi di Perugia  and Istituto  Nazionale di 
Fisica Nucleare, Sezione di Perugia, Via A. Pascoli, 00610 Perugia, Italy}

\noindent { $^d$ \it Istituto Nazionale di Fisica Nucleare, Sezione di Pisa,  
 Via
Buonarroti 2, 56100 Pisa, Italy}

\vspace{1.5cm}

\begin{abstract}  
The possibility of a reliable extraction of the neutron deep inelastic structure
function, $F_2^n(x)$, for  $ x < 0.85$ from joint measurements of
deep inelastic structure functions of deuteron, $^{3}He$ and $^{3}H$ is
investigated. The model dependence in this extraction, linked to the possible
different interactions between nucleons in nuclei, is shown to be weak, if 
the nuclear structure effects are properly taken into account.
A combined analysis of the deep inelastic structure functions of these nuclei is
proposed to study effects beyond the impulse approximation. 
\end{abstract}

\vspace{4.5cm}

\hrule width5cm
\vspace{.2cm}
\noindent{\normalsize{ To appear in 
{\bf Phys. Rev. C}.}}

\newpage
\pagestyle{plain}

\section{INTRODUCTION}
\indent 
The knowledge of both proton and neutron deep inelastic structure functions
(DISF's) at large values of the Bjorken variable $x = Q^2 / (2 M \nu)$ could give
access to the valence $u$ and $d$ quark distributions in the nucleon \cite{ud}
($Q^2=-q^2$, with $q \equiv(\nu, \vec{q})$  the four-momentum transfer).
Usually, deuteron data have been employed to gain information on the neutron
unpolarized DISF, $F_2^n(x)$, but uncertainties remain, linked to the EMC effect
in the deuteron \cite{Frank,Kaptari}. Medium off-shell effects, different from
the binding effects and related to a nucleon structure in nuclei different from
the free one, have been often advocated (see, e.g., Refs. \cite{Gross,Umni}).
Although these effects have been found to be small, it was argued
\cite{thom} that the standard treatment of deuteron data \cite{Bodek} could be
unfair. Indeed, such a treatment employes convolution formulas, neglecting
medium effects beyond the impulse approximation (IA), and leads to the value
$1/4$ for the ratio  $r(x) = F_2^n(x)/F_2^p(x)$, when $x \rightarrow 1$. At
variance, an analysis which includes medium effects \cite{thom} moves such a
value towards that of $3/7$, suggested by pQCD arguments \cite{fj}.

Recently, the possible use of an unpolarized $^3H$ target has been discussed
\cite{JCR}. In particular, an experiment has been proposed \cite{MP,Pet}, aimed
at determining $F_2^n(x)$ at large $x$ from the measurement of the ratio
$E^{HeT}(x) = F_2^{He}(x)/F_2^T(x)$ between the unpolarized structure functions of
$^{3}He$, $F_2^{He}(x)$, and $^{3}H$, $F_2^T(x)$. Indeed, using this ratio one is
expected to reduce the effects of systematic errors in the measurements, as well
as the effects of theoretical model dependences and, in particular, of
contributions beyond the impulse approximation. As far as the latter are
concerned, the differences between the EMC effect in $^{3}He$ \cite{Ack} and in
$^3H$ are expected to be small, because of isospin symmetry \cite{MP}.

In Ref. \cite{PSS} a reliable recurrence procedure has been proposed, within the
impulse approximation \cite{CL,CSPS}, to extract $F_2^n(x)$ in the range $0 < x
\le 0.9$ from the experimental ratio $E^{HeT}(x)$. It has been shown that, at high
$x$, nuclear structure effects, i.e., Fermi motion and nuclear binding, are
relevant and cannot be overlooked. 

In this paper, using the same approach of Ref. \cite{PSS}, we will show that 
the extraction of $F_2^n(x)$ from the ratio $E^{HeT}(x)$, up to $x < 0.85$, is weakly  
dependent upon 
the
different possible interactions between nucleons in nuclei. Furthermore, we
suggest a method to check the role of effects beyond the impulse approximation
and the reliability of the many different expressions proposed for the description
of the DISF's of nuclei (see, e.g., Refs. \cite{Frank,Kaptari,CL,FS,Coe}).
 Our approach is based on
 a joint analysis of the experimental ratios of: i) deuteron to proton,
$E^{Dp}(x) = F_2^D(x)/F_2^p(x)$, ii) $^{3}He$ to deuteron, $E^{HeD}(x) = 
F_2^{He}(x)/F_2^D(x)$, and iii) $^{3}He$ to $^3H$, $E^{HeT}(x) = F_2^{He}(x)/F_2^T(x)$,
DISF's.
 We accurately take care of nucleon motion and nucleon binding in the two- and
three-nucleon systems and explicitly consider the Coulomb interaction in the
evaluation of the $^{3}He$ spectral functions. To this end, we take advantage
of the very accurate wave functions of $^{3}He$ and $^3H$ systems, which can be
calculated for realistic interactions within the correlated
hyperspherical harmonics (CHH) approach of Ref. \cite{KRV}.
  For an easy presentation, only the case of
infinite momentum transfer in the Bjorken limit is considered, but it is
straightforward to generalize our approach to the realistic case of finite
momentum transfer values \cite{PSS1}.

The paper is organized as follows : in Sec. II, the general formalism for the
DISF's is presented; in Sections III and IV recurrence relations for the extraction
of  $F^n_2(x)$ from DISF's of
few-nucleon systems are proposed and the sensitivity to the interaction between
nucleons  is investigated; conclusions are drawn in Sec. V.

\section{GENERAL FORMALISM}
\indent Our analysis is based on IA, which is usually employed for the calculation of
nuclear structure functions at intermediate values of $x$, i.e., when the very
small-$x$ and the very large-$x$ regions are excluded \cite{Gee}. In IA
the nucleon structure is assumed to be the same as for free nucleons, and the
DISF's for the deuteron, $F_2^D(x)$, for $^{3}He$, $F_2^{He}(x)$, and for $^{3}H$,
$F_2^T(x)$ can be written, in the Bjorken limit, as follows:  
\begin{eqnarray}
F_2^{D}(x) = 
\int\nolimits_{x}^{M_D/M} [ F_2^{p}(x/z) + F_2^{n}(x/z) ] f^{D}(z) dz  
\label{33}
\end{eqnarray}
\begin{eqnarray} 
&& F_2^{He}(x) = 
2 \int\nolimits_{x}^{M_{He}/M}F_2^{p}(x/z) f_{p}^{He}(z) dz + 
\int\nolimits_{x}^{M_{He}/M} F_2^{n}(x/z) f_{n}^{He}(z) dz 
\label{34}
\end{eqnarray}
\begin{eqnarray}  
&&F_2^{T}(x) = \int\nolimits_{x}^{M_T/M} F_2^{p}(x/z) f_{p}^{T}(z) dz+
2 \int\nolimits_{x}^{M_T/M} F_2^{n}(x/z) f_{n}^{T}(z) dz  
\label{35}
\end{eqnarray}
where $M$, $M_D$, $M_{He}$, $M_T$ are the masses of nucleon, deuteron, $^{3}He$ and $^3H$,
respectively. Different expressions have been proposed, see e.g. Refs. \cite
{Kaptari,Coe}, for the distributions $f^D(z)$ and $f_{p(n)}^{He(T)}(z)$, which 
 describe
the structure of the deuteron and of the three-nucleon systems. In this paper,
we consider the following ones  
\cite{CL} 
\begin{eqnarray} 
f^{D}(z) = 
\int\nolimits d {\vec p} \; n^D( |{\vec p} |) \; \delta \left( z -
\frac {pq} {M\nu} \right) \; z \; C \;  
\label{2}
\end{eqnarray}
\begin{eqnarray}
&& f_{p(n)}^{He(T)}(z) = 
\int\nolimits d E 
\int\nolimits d {\vec p} \; P^{He(T)}_{p(n)}( |{\vec p} |, E) \; \delta \left( z -
\frac {pq} {M\nu} \right) \; z \; C' \; . 
\label{1}
\end{eqnarray}

In Eqs. (\ref{2}) and (\ref{1}), $n^D( |{\vec p} |)$ is the nucleon momentum
distribution in deuteron, the functions $P^{He(T)}_{p}(|{\vec p}|, E)$ and
$P^{He(T)}_{n}(|{\vec p}|, E)$ are the proton and neutron spectral functions 
in $^{3}He$ ($^{3}H)$, respectively, \cite{CPS} ${\vec p}$ and $E$ the nucleon
three-momentum and removal energy, $C$ and $C'$ normalization factors.  
The Coulomb interaction is explicitly taken into account in the evaluation of
the $^{3}He$ spectral function, unless otherwise explicitly specified. In Eqs.
(\ref{2}) and (\ref{1}), to ensure the 4-momentum conservation at the virtual
photon-nucleon vertex, the nucleon is assumed to be off-mass shell, i.e.,  $p \equiv
(p^0, {\vec p})$ with $p^0 = M_{D(He,T)} -  \sqrt{(E + M_{D(He,T)} - M)^2 + |{\vec p}
|^2}$.

It has to be noted that the definitions (\ref{2}) and (\ref{1}) of the
distributions $f^D(z)$ and $f_{p(n)}^{He(T)}(z)$, because of the off-mass-shell
nucleon energy, $p^0$, already include the off-shell effects considered in the
$x-$rescaling model of Ref. \cite{Akul} (and, therefore, also the effects
related to the derivative of the nucleon structure functions studied in Ref.
\cite{Burov}).    

\section{EXTRACTION of $F^n_2(x)$ from $^{3}He$ and $^{3}H$ DISF's}
\indent To perform our study we use the proton and neutron spectral functions for
$^{3}He$ and $^{3}H$ that were obtained in Ref. \cite{KPSV} with the $RSC$ \cite{RSC},
$AV14$ \cite{AV14}, and $AV14$ + Brazil three-body force (TBF) \cite{Br} interactions 
(in the last case the Brazil three-body force was properly tuned in 
Ref. \cite{KRV} to obtain the experimental binding energy of $^3H$).
   Furthermore, we have specifically evaluated, along the same lines of Ref.
\cite{KPSV}, the spectral functions for the $AV18$ interaction \cite{AV18}, for the
$AV18$ + $Urbana IX$ ($UIX$) TBF interaction \cite{UIX} and for the $AV18$ + $TM'$ TBF
interaction, which is a new version of the original $Tucson-Melbourne$ ($TM$) TBF
\cite{TM}, from the corresponding CHH wave functions. Note that the $UIX$ TBF was
specifically proposed in Ref. \cite{UIX} to get, together with the $AV18$ two-body
interaction, the experimental binding energy of light nuclei and reproduces the
binding energies of both $^{3}He$ and $^{3}H$. The $TM'$ TBF was properly modified 
in Ref. \cite{Friar} to
ensure consistency with chiral symmetry. The values of the strength and
cutoff parameters of the $TM'$ TBF are taken from Ref. \cite{Witala}: with these
values the $AV18$ + $TM'$ interaction describes the $A = 3$ ground state energies.

 Let us define the super-ratio,  $R^{HeT}(x)$ \cite{MP},
\begin{eqnarray}
&& R^{HeT}(x) = \frac{F_2^{He}(x)/(2F_2^p(x) 
+ F_2^n(x))}{F_2^T(x)/(2F_2^n(x) + F_2^p(x))} = \nonumber \\ 
&& E^{HeT}(x) \; \; \frac{2 r(x) + 1} {2 + r(x)} \; \;. 
\label{37}
\end{eqnarray} 

In IA the super-ratio is a functional of $r(x)$ ( $R^{HeT}(x) = R^{HeT}[x,r(x)]$ ).
Indeed from Eqs. (\ref{34},\ref{35},\ref{37}) one has
\begin{eqnarray}
&& R^{HeT}[x,r(x)] = \frac{2 r(x) + 1} {2 + r(x)}\nonumber \\ 
&& \frac{ \int\nolimits_{x}^{M_{He}/M} F_2^p(x/z)~ \left [ 2f_{p}^{He}(z)   + 
 r(x/z)  f_{n}^{He}(z) \right ]~ dz}
{\int\nolimits_{x}^{M_T/M} F_2^p(x/z)~\left [ f_{p}^{T}(z)  + 
2  r(x/z)  f_{n}^{T}(z) \right ]~ dz  } \; \; .
\label{38}
\end{eqnarray} 
The extraction of $r(x)$ from the experimental ratio $E^{HeT}(x)$ can proceed,
through Eq. (\ref{37}), with the help of theoretical estimates of $R^{HeT}(x)$. 
Actually, from Eq. (\ref{37}) one can immediately obtain the following equation 
for the ratio $r(x)$ :  
\begin{eqnarray} 
r(x) = \frac{E^{HeT}(x) - 2 R^{HeT}[x,r(x)]}{R^{HeT}[x,r(x)] - 2E^{HeT}(x)} 
\label{36}
\end{eqnarray}

In IA, Eq. (\ref{36}) is a self-consistent equation, which allows one to 
determine $r(x)$. If the distributions $f_{p(n)}^{He(T)}(z)$ are represented by a
Dirac $\delta$ function, $f_{p(n)}^{He(T)}(z) = \delta (z-1)$, then
$R^{HeT}(x) = 1$ and Eq. (\ref{36}) becomes trivial. This hypothesis works
reasonably well at small $x$, but is not a good approximation at $x > 0.75$, so
that $R^{HeT}(x) \neq 1$, as shown in Fig. 1. As a consequence, if: i) the
experimental quantity $E^{HeT}(x)$ is simulated by its theoretical estimate,
evaluated in IA through Eqs. (\ref{34}) and (\ref{35}) with some model for
$r(x)$;  and ii) the approximation  $R^{HeT}(x) = 1$ is used to calculate $r(x)$
by Eq. (\ref{36}), then one obtains a function which differs  $\sim 10 \%$ at
$x = 0.85$ from the model for $r(x)$ used to simulate $E^{HeT}(x)$. Therefore,
at high $x$ one cannot approximate $R^{HeT}(x)$ by 1, if a good accuracy is
required.

Fortunately, as illustrated in Fig. 1(a), the model dependence of $R^{HeT}(x)$,
due to the different, possible two-body and three-body interactions between
nucleons in $^{3}He$ and $^{3}H$, is very weak for any $x$. Indeed, there is a
substantial cancellation of interaction effects in the numerator and in the
denominator. In particular, the introduction of a three-body force yields
very small effects in $R^{HeT}(x)$ at $x \le 0.90$. Only if the
Coulomb interaction is neglected in the evaluation of the $^{3}He$ spectral
functions, one obtains relevant effects, since this interaction acts exclusively
in the numerator. However, sensible differences in $R^{HeT}(x)$ are obtained for
$x \ge 0.5$, if the nucleon spectral functions are replaced by the corresponding
nucleon momentum distributions, since this approximation yields much larger
effects in $f_{p(n)}^{He(T)}(z)$ than the different interactions (see also
Ref. \cite{CL} for the relevance of the spectral function). 

Since several  models for the nucleon DISF's, to be used in
Eq. (\ref{38}),
are available, the sensitivity of the super-ratio to
the different parametrizations has also to be checked.
A scale of $Q^2$= 10 $(GeV/c)^2$ has been chosen
for the evaluation of the nucleon DISF's. Such a scale
is low enough to allow the use of many of the
available models and, at the same
time, relevant differences are not expected
between the results obtained at $Q^2$= 10 $(GeV/c)^2$
and the ones corresponding to the Bjorken limit.
The super-ratio Eq. (\ref{38}), evaluated
by using the DISF's given in \cite{Aubert}, \cite{MRST}, \cite{DL}
and \cite{SMC},
is shown in Fig. 1 (b).
 
In Ref. \cite{PSS} we showed that, within IA, Eq. (\ref{36}) can be solved by
recurrence     
\begin{eqnarray}  
r^{(n+1)}(x) = \frac{E^{HeT}(x) - 2 \, R^{HeT}[x,r^{(n)}(x)]}
{R^{HeT}[x,r^{(n)}(x)] - 2 \, E^{HeT}(x)}  
\label{39}
\end{eqnarray}
starting from a reasonable zero-order approximation, $r^{(0)}(x)$. Since no data
are presently available for $F_2^T(x)$, we simulated the experimental ratio
$E^{HeT}(x)$ by a theoretical IA estimate. Both $E^{HeT}(x)$ and $R^{HeT}(x)$ were
evaluated with the same nucleon spectral functions. The nucleon DISF's of Ref.
\cite{Aubert} were used in the calculation of $E^{HeT}(x)$, while, to generate the
zero-order approximation $r^{(0)}(x)$ to be used in $R^{HeT}(x)$, the neutron
one was arbitrarily modified by the factor $(1 + 0.5x^2)$ to change its
behaviour at high $x$. Using the nucleon spectral functions obtained from the $AV18$ 
+ $UIX$ TBF interaction, a sequence which rapidly converges to $r(x)$ of
Ref. \cite{Aubert} is obtained in the range $0 \le x \le 0.9$. In particular,
up to $x = 0.85$ an accuracy better than $1 \%$ is obtained with only ten
iterations.  Starting from very different zero-order  approximations
$r^{(0)}(x)$, for instance the ratios corresponding to the nucleon DISF's of
Refs. \cite{MRST}, \cite{DL} or \cite{SMC} (see Fig. 2(a)), while still
evaluating $E^{HeT}(x)$ from the nucleon DISF's of Ref. \cite{Aubert},
convergences of a similar quality and to the same $r(x)$ of Ref. \cite{Aubert}
have  been obtained. Therefore, one can conclude that, up to $x = 0.85$, the
recurrence relation converges to the correct result, almost independently of the
starting point $r^{(0)}(x)$ (see the dot-dashed line in Fig. 2(b)). A convergence
of the same quality is obtained if the spectral functions used for the
calculation of $E^{HeT}(x)$ and $R^{HeT}(x)$ correspond to another interaction, e.g.
the RSC interaction \cite{PSS}.
 The convergence of the recurrence relation to the correct result
can be related to the similarity between the distributions $f_{p(n)}^{He(T)}(z)$ and
$\delta (z-1)$. Near $x \sim 1$, where  $f_{p(n)}^{He(T)}(z)$ no more acts as a Dirac
$\delta$ function in Eqs. (\ref{34}) and (\ref{35}), the recurrence relation is unable
to solve Eq. (\ref{36}).     

In order to check the model dependence of our approach, due to the different
assumptions for the interaction between nucleons in nuclei,  we repeat the whole
procedure of Ref. \cite{PSS},  but using for the evaluation of the super-ratio
$R^{HeT}(x)$ spectral functions corresponding to different interactions than the
$AV18$ + $UIX$ one, employed for the calculation of our simulated "experimental" ratio
$E^{HeT}(x)$. The spectral functions corresponding to $RSC$, $AV14$, $AV14$
+ Brazil TBF, $AV18$ and $AV18$ +  $TM'$ TBF 
interactions are considered.
 In the range $0 \le x \le 0.85$ the ratio $r(x)$ extracted by the recurrence
relation after twenty iterations differs from the one used for $E^{HeT}(x)$ less
than $3 \%$, for any of the considered interactions (see Fig. 2 (b)). Actually, if
only interactions able to give the experimental value for the binding energy of
$^3H$ are considered (i.e., $AV14$ + Brazil, $AV18$ + $TM'$ and $AV18$ + $UIX$),
the model dependence in the extraction of $r(x)$ is at most $1\%$ in the range 
$0 \le x \le 0.85$.
 Furthermore, these results are essentially
independent of the model for the ratio $r(x)$, which is used in the evaluation of
$E^{HeT}(x)$. Let us stress that the recurrence procedure yields somewhat larger
differences with respect to the input $r(x)$ ($\sim 4\%$ at $x = 0.85$), if the
Coulomb interaction is neglected in the $^3He$ spectral functions considered for the
evaluation of $R^{HeT}(x)$. However, these differences are not to be included in the
model dependences, since the Coulomb interaction can be exactly taken into account,
e.g. within the CHH approach.

Let us note that, in order to apply the recurrence relation (\ref{39}), the
knowledge of the function $E^{HeT}(x)$ is needed on the whole range $0 < x < 1$,
even if one is interested in $r(x)$ for $x < 0.85$ only.  However, in the near
future $E^{HeT}(x)$ will not be experimentally accessible for $x \ge 0.85$.  
To investigate the possible effects on the extraction of $r(x)$ due to
this problem, we
change the "experimental" ratio $E^{HeT}(x)$ by an arbitrary factor $(1 + 0.5 x^{20})$,
which
modifies only the large $x$ region, and repeat the recurrence extraction
procedure. The ratio $E^{HeT}(x)$ is essentially
unchanged by the factor $(1 + 0.5 x^{20})$
up to $x=0.8$, is modified by $2 \%$ at $x = 0.85$ and by 
$50 \%$ at $x = 1$. Then, after twenty iterations one obtains convergence to the
same $r(x)$ up to $x = 0.8$ and only a $5 \%$ difference at $x = 0.85$.

Therefore, within IA, the proposed procedure is able to yield reliable
information on $F_2^n(x)$ in the $x$ range accessible at TJLAB \cite{MP},
whenever nucleon binding in nuclei and the Coulomb interaction in the $^{3}He$
spectral function are correctly taken into account. On the contrary, if the
momentum distribution is used for the evaluation of $R^{HeT}(x)$, instead of the
spectral function, the iterative procedure converges to a function $r(x)$, which
differs from the correct one more than $13 \%$ for $x \ge 0.8$ (see Fig. 2(b)).

Let us note that our results hold unchanged if, instead of 
Eq. (\ref{1}), a different expression (see, e.g., \cite{Kaptari,FS,Coe}) is used to
evaluate both $E^{HeT}(x)$ and $R^{HeT}(x)$. 

 \section{EXTRACTION of $F^n_2(x)$ from a joint analysis of $^2H$, $^{3}He$ and $^{3}H$
DISF's}

Many different expressions have been proposed to describe the DISF's of nuclei 
 and to explain the EMC effect (see, e.g., Refs. \cite{Frank,Kaptari,CL,FS,Coe} and
references quoted therein), which are based on convolution formulas or involve medium
effects beyond IA. The different models can clearly affect the extraction of
$F^n_2(x)$.
 For instance, in Ref. \cite{thom} it was shown that, at large $x$, medium effects
beyond IA can considerably modify the neutron DISF extracted from the
experimental deuteron DISF. Again in the case of the deuteron, in Ref. \cite{PS} it
was shown that sizeable effects are obtained in the extraction of neutron DISF
if, instead of the model given by Eqs. (\ref{33}) and (\ref{2}), one adopts a
convolution model developed within the front-form Hamiltonian dynamics with a
Poincar\'e-covariant current operator. 
Although the effects of the different
expressions proposed for the DISF's of nuclei are present
both in the numerator and in the denominator of $R^{HeT}(x)$ and they should at least
partially compensate in the ratio, their relevance in the extraction of $F_2^n(x)$ has
to be carefully investigated. A possible check of the correctness of the different
theoretical expressions could be performed by comparing the neutron DISF,
independently extracted from the experimental ratios $E^{Dp}(x)$, $E^{HeD}(x)$ and
$E^{HeT}(x)$, using a coherent framework for the evaluation of the deuteron, $^3He$ and
$^3H$ DISF's. Indeed, the theoretical super-ratios corresponding to $E^{Dp}(x)$
and $E^{HeD}(x)$ will be much more affected by the model used for the evaluation
of the structure functions than in the case of $R^{HeT}(x)$. Then, one can take
advantage of this model dependence for a test of the  theoretical models:  the proper expressions 
for the DISF's of nuclei should lead
to the same results for the neutron DISF extracted from any of the above mentioned
experimental ratios. 

 This analysis, performed with actual estimates of medium correction terms or using
different convolution formulas, is outside the scope of the present work. Here we only
wish to show that $ F_2^n(x)$ can be extracted from the ratios $E^{Dp}(x)$ and
$E^{HeD}(x)$ through the following recurrence relations, based on IA and analogous to
the one of Eq. (\ref{39}),  
\begin{eqnarray}  
&& r^{(n+1)}(x) = \frac{E^{Dp}(x) } {R^{Dp}[x,r^{(n)}(x)] }-1 = \nonumber \\
&&\frac{F_2^{D exp}(x) [1 + r^{(n)}(x) ] } 
{\int\nolimits_{x}^{M_D/M} [ 1 + r^{(n)}(x/z) ] F_2^{p}(x/z) f^{D}(z) dz } - 1 
\label{39p} 
\end{eqnarray}
\begin{eqnarray}  
r^{(n+1)}(x) = \frac{E^{HeD}(x) - 2 \, R^{HeD}[x,r^{(n)}(x)]}
{R^{HeD}[x,r^{(n)}(x)] - E^{HeD}(x)}  
\label{39s}
\end{eqnarray}
with natural definitions for the super-ratios 
$R^{Dp}(x) = F_2^D(x)/[ F_2^p(x)+ F_2^n(x)]$ and 
$R^{HeD}(x) = F_2^{He}(x) [ F_2^p(x) + F_2^n(x)] / \{ F_2^D(x) [2 F_2^p(x) +
F_2^n(x)] \} $.
 
As we did before for Eq. (\ref{39}), we simulate the experimental ratios 
$E^{Dp}(x)$ and $E^{HeD}(x)$ by theoretical estimates using Eqs.
(\ref{33},\ref{34},\ref{2},\ref{1}) with a given momentum distribution or
given spectral functions for the deuteron and for $^3He$, respectively . Then, we
evaluate $R^{Dp}(x)$ and $R^{HeD}(x)$ in IA with the same
nucleon momentum distribution or spectral functions, and assume a function
$r^{(0)}(x)$ as the zero-order approximation. As shown in Fig. 3, the convergence of
the recurrence relations (\ref{39p}) and (\ref{39s}) to the input model for the ratio
$r(x)$, used for the calculation of the simulated "experimental" quantities, is very
fast. Furthermore, as in the case of Eq. (\ref{39}), the extracted $r(x)$ is
essentially independent of the function $r^{(0)}(x)$, assumed as the zero-order
approximation.

  The evaluation of the model dependence due to the nuclear interaction in the
extraction of $r(x)$ by the recurrence relations (\ref{39p}) and (\ref{39s}) deserves
a separate analysis for each one of these two equations. For the deuteron-proton case
(Eq. (\ref{39p})) we evaluate $R^{Dp}(x)$ by means of different nucleon momentum
distributions than the one corresponding to the $AV18$ interaction used to simulate
$E^{Dp}(x)$. Using any of the already mentioned two-body interactions, the function
$r^{(n)}(x)$ obtained after twenty iterations differs less than $4 \%$ up to $x \le
0.80$ and by $8 \%$  at $x = 0.85$ from the function $r(x)$ used for  $E^{Dp}(x)$.
For the $^3He$-deuteron case (Eq. (\ref{39s})), we simulate $E^{HeD}(x)$
through Eqs. (\ref{33}) and (\ref{34}), using the  $AV18$ interaction for the deuteron
and the $AV18$ + $UIX$ TBF interaction for the $^3He$, respectively. Then, using any of
the mentioned two-body and three-body interactions to evaluate $R^{HeD}(x)$,
differences as high as $12 \%$ at $x = 0.8$ and $25 \%$ at $x = 0.85$ are found
between $r^{(20)}(x)$ and the parametrization for $r(x)$ used to simulate $E^{HeD}(x)$.
However, if the model dependence in the extraction of $r(x)$ due to the nuclear
interaction is estimated, as it has to be, considering only the differences generated
by interactions able to correctly reproduce the experimental binding energy of $^3H$,
then the effects of the possible different interactions is reduced to  $1 \%$ at most,
up to $x = 0.85$. Therefore, both for $E^{Dp}(x)$ and $E^{HeD}(x)$, the effects of
the different nuclear interactions  on the extraction of
$F_2^n(x)$ are well under control.

\section{CONCLUSIONS}
\indent In this paper, recurrence relations for the extraction of $F^n_2(x)$
for $x<0.85$ from DISF's of deuteron, $^3He$ and $^3H$ have been proposed within IA.
These recurrence relations, which require a zero-order approximation for the neutron
structure function $F^n_2(x)$, have been shown to be rapidly convergent and
essentially insensitive to the zero-order approximation. Moreover, they are only very
weakly dependent on the interaction between nucleons in nuclei, whenever the A = 3
binding energies are correctly reproduced. In the case of the three-nucleon systems,
the relevance of accurate calculations which
take into account the nuclear structure by means of
 the spectral function was stressed. In particular, we have investigated 
the role played by the Coulomb interaction in  $^3He$, for a good accuracy in the
extraction of $F^n_2(x)$ at high values of $x$.

   Summarizing, we suggest to take advantage of the very well
 known nuclear
structure of few-nucleon systems to extract $F_2^n(x)$ from a joint analysis of
deuteron, $^3He$ and $^3H$ DISF's. Our approach can be easily extended to include
the analysis of $^4He$ DISF. We stress that, while waiting for the $^3H$ experiments
in order to perform a more complete investigation, a simultaneous analysis of the
experimental ratios $E^{Dp}(x)$ and $E^{HeD}(x)$ in a wide range of $x$ should be
carried out. In these cases, the model dependences in the evaluation of the structure
functions will be bigger than in the  $E^{HeT}(x)$ case, but the comparison of the 
results obtained from the
recurrence relations (\ref{39p}) and (\ref{39s}), including possible contributions
beyond IA in the evaluation of the super-ratios $R^{Dp}(x)$ and $R^{HeD}(x)$, could
already give useful information on the role of medium effects and consequently allow a
more reliable extraction of $F_2^n(x)$.  In the case of $E^{HeD}(x)$, one should
accurately take care of three-body forces which give the experimental $^3H$ binding
energy. Indeed considerable differences are obtained if interactions which do not
reproduce $^3H$ binding energy are used in the evaluation of $R^{HeD}(x)$. At variance,
because of isospin symmetry, these effects largely cancel out in the ratio of $^3He$
to $^3H$ DISF. This fact supports the usefulness of measurements of the ratio
$E^{HeT}(x)$ for the extraction of the neutron deep inelastic structure function.

\section{ACKNOWLEDGMENTS}

We would like to thank  L. P. Kaptari, W. Melnitchouk and G.G. Petratos for many 
helpful
discussions and A. Molotchkov for providing a computer routine for the SMC
parametrization of DISF's and A. Donnachie for providing the computer routine
for DISF model of Ref. \cite{DL}. One of the authors (S.S.) thanks S. Simula for useful
discussions during the workshop "Hix2000", held in Philadelphia in April 2000.
This work was partially supported by the Italian Ministero dell'Universit\`a e
della Ricerca Scientifica e Tecnologica.

\newpage
\onecolumn
\begin{center}
FIGURE CAPTIONS
\end{center}
Figure 1.{ (a) The super-ratio $R^{HeT}(x)$ (Eq. (\ref{38})) with $F_2^{n(p)}(x)$
from Ref. \cite{Aubert} for different nuclear interactions. Solid and 
long-dashed lines correspond to the $AV18$ + $UIX$ TBF and $RSC$ interactions,
respectively (the results for $AV18$, $AV18$ + $TM'$ TBF, $AV14$, and $AV14$ + Brazil
TBF are essentially identical to the ones for $AV18$ + $UIX$ TBF and are not
shown). The short-dashed line corresponds to the $AV14$ interaction without the
Coulomb interaction for $^{3}He$. The dotted line is obtained as the solid one, but
using the nucleon momentum distributions for the $AV18$ + $UIX$ interaction, instead of the
nucleon spectral functions. (b) The super-ratio $R^{HeT}(x)$ for the $AV18$ + $UIX$
 interaction.
Dashed, dotted, and solid lines correspond to the models of Refs. \cite{MRST}, 
\cite{DL} and  \cite{Aubert} for $F_2^{n(p)}(x)$, respectively (the
model of Ref. \cite{SMC} gives almost identical results of the model of Ref.
\cite{DL} and is not shown); long-dashed line: as the solid one with $F_2^{n}(x)$
multiplied by $(1 + 0.5 x^2)$ (see text). }

\vspace {1cm}

Figure 2. { (a) The ratio $r(x)$ for different parametrizations of nucleon
DISF's. Thick-solid, dashed,  dotted, and thin-solid lines correspond to the models
of Refs. \cite{Aubert}, \cite{MRST}, \cite{DL},
and \cite{SMC} for the nucleon DISF's, respectively. The long-dashed line
corresponds to the  model of Ref. \cite{Aubert}, multiplied by $(1 + 0.5 x^2)$.
(b) $r^{(n)}(x)$, obtained by the recurrence relation (\ref{39}) for $n = 20$,
using the nucleon DISF's of Ref. \cite{Aubert} and the $AV18$ + $UIX$
 spectral function
for $E^{HeT}(x)$.  Different spectral functions are used for $R^{HeT}(x)$:
dot-dashed, thin-solid, dashed and long-dashed lines correspond to the $AV18$ + $UIX$
TBF, $AV18$, $AV14$ and $RSC$ spectral functions,
respectively (the results for $AV14$ + Brazil TBF and for $AV18$ + $TM'$ TBF are
almost indistinguishable up to $x = 0.9$ from the ones for $AV18$ + $UIX$ TBF and are
not shown). The dotted line is $r^{(20)}(x)$, obtained using the nucleon momentum
distributions for the $AV18$ + $UIX$ TBF interaction in the evaluation of $R^{HeT}(x)$,
instead of the spectral functions. The thick solid line is the ratio $r(x)$ for the
nucleon DISF's of Ref. \cite{Aubert}. }

\vspace {1cm}

Figure 3. { (a) The ratio $r(x)$ obtained by the recurrence relation
(\ref{39p}), using the $AV18$ interaction both for $E^{Dp}(x)$ and $R^{Dp}(x)$.
Long-dashed, thin-solid and dot-dashed lines are $r^{(n)}(x)$ for $n = 3, 6, 20$
iterations, respectively. The thick solid line is the ratio $r(x)$ for the
nucleon DISF's of Ref. \cite{Aubert}, used to evaluate $E^{Dp}(x)$.
(b) The same as in (a), but for the recurrence relation (\ref{39s}), concerning
the ratio of $^3He$ to deuteron DISF. The $AV18$ and the $AV18$ + $UIX$ TBF 
interactions
have been used for the $^2H$ and $^3He$ DISF's, respectively. }

\newpage
\begin{figure}[t]

\psfig{figure=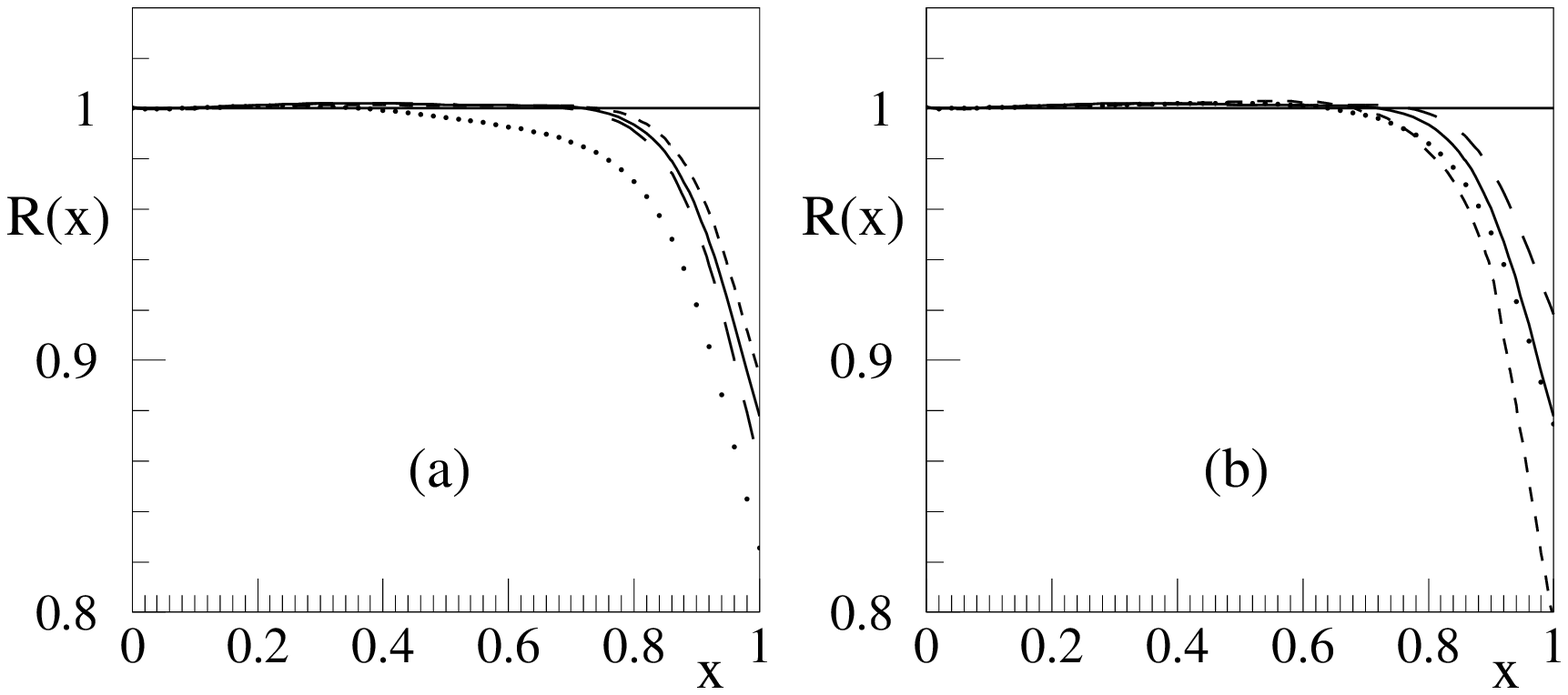,bbllx=25mm,bblly=180mm,bburx=10mm,bbury=280mm}  

\vspace {3cm} 
Fig. 1   E. PACE, G. SALM\`E, S. SCOPETTA, A. KIEVSKY

\end{figure}

\begin{figure}[t]

\psfig{figure=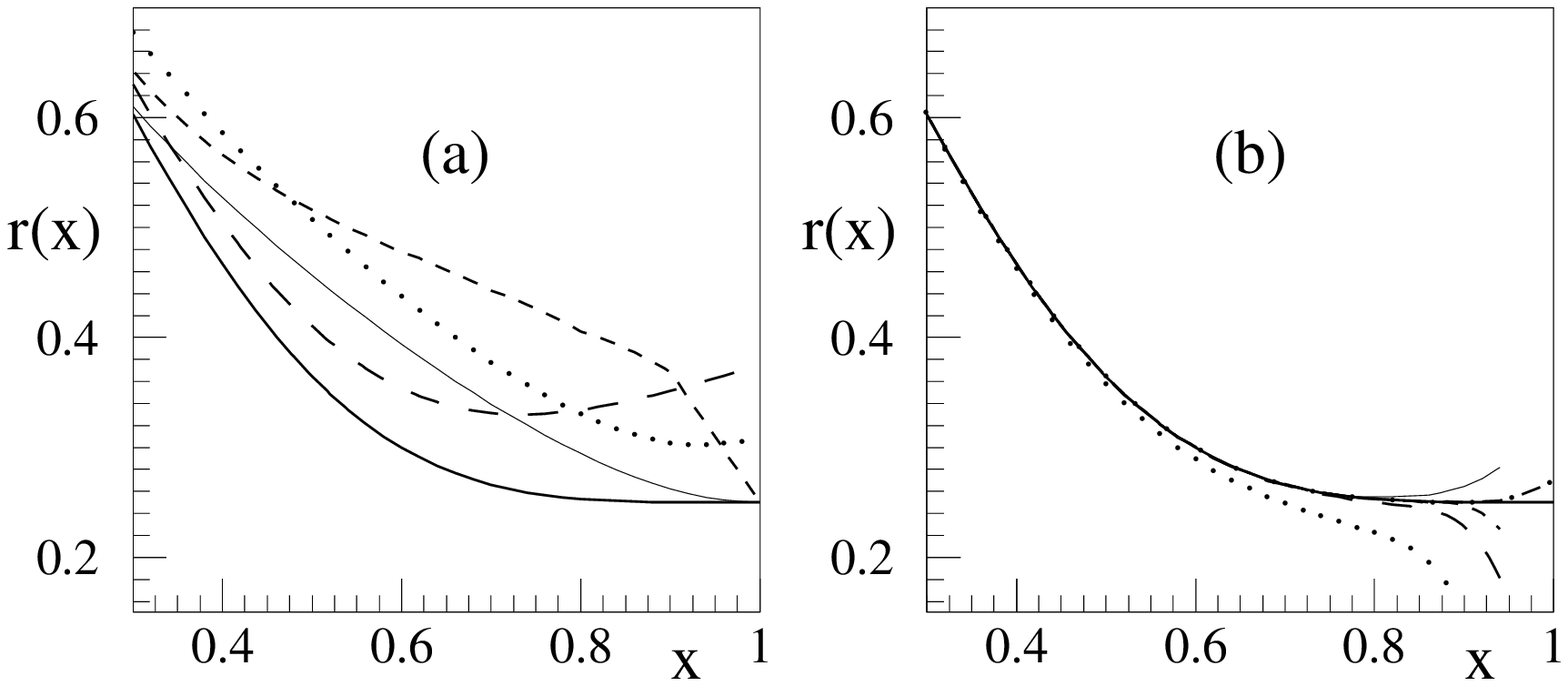,bbllx=25mm,bblly=180mm,bburx=10mm,bbury=295mm}  

\vspace {3cm} 
Fig. 2   E. PACE, G. SALM\`E, S. SCOPETTA, A. KIEVSKY                
\end{figure}

\begin{figure}[t]

\psfig{figure=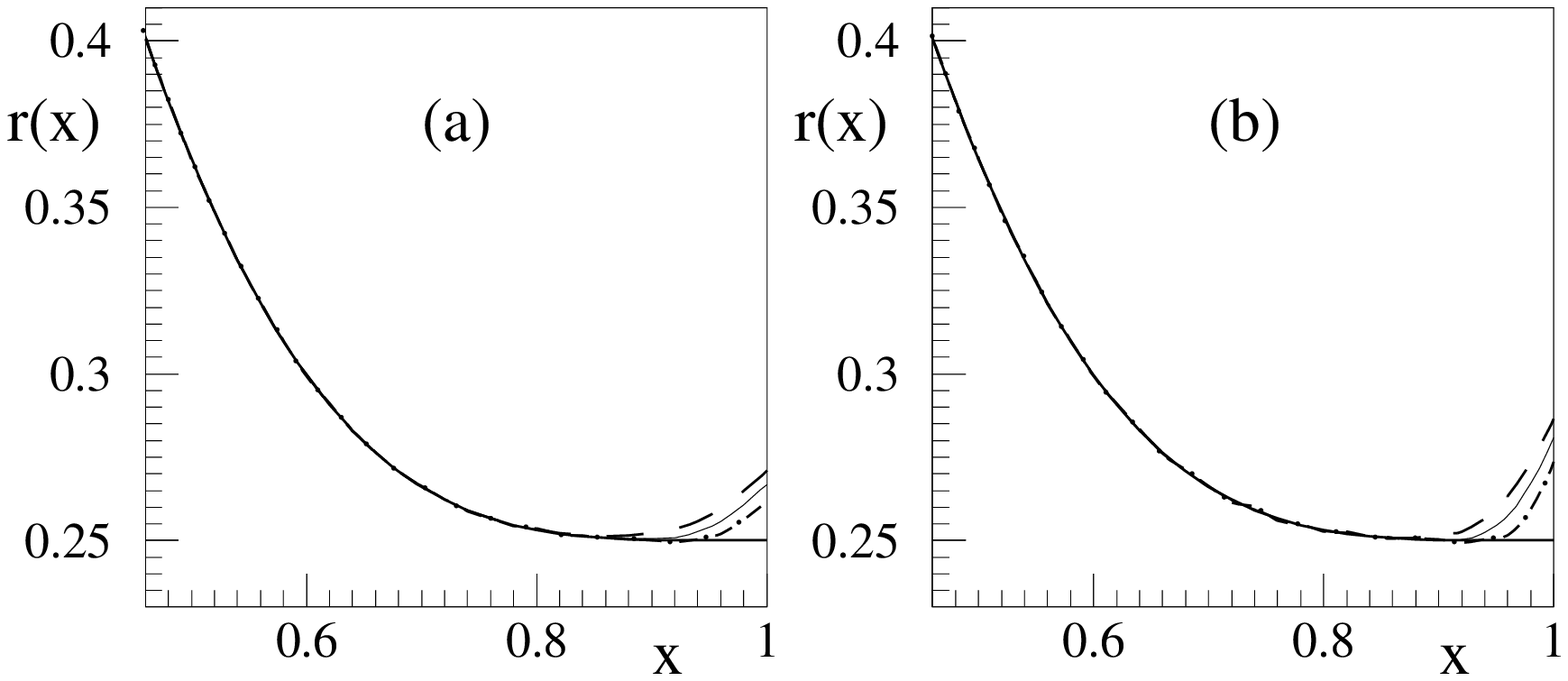,bbllx=25mm,bblly=180mm,bburx=10mm,bbury=295mm}  
 
 \vspace {3cm} 
Fig. 3   E. PACE, G. SALM\`E, S. SCOPETTA, A. KIEVSKY     
\end{figure} 

\end{document}